\documentstyle[epsf,epsf,aps]{revtex}

\begin{document}

\title{Glassy dynamics in granular compaction: sand on random graphs}

\author{Johannes Berg$^1$ and Anita Mehta$^2$}

\address{1 - Abdus Salam International Centre for Theoretical Physics, 
34100 Trieste, Italy}
\address{2 - S N Bose National Centre for Basic Sciences, Block JD
Sector III, Salt Lake, Calcutta 700 098, India}

\maketitle
 
\begin{abstract}
We discuss the use of a ferromagnetic spin model on a random graph to 
model granular compaction. A multi-spin interaction is used to capture the 
competition between local and global satisfaction of constraints 
characteristic for geometric frustration. We define an athermal dynamics 
designed to 
model repeated taps of a given strength.  
Amplitude cycling and the effect of 
permanently constraining a subset of the spins at a given amplitude is 
discussed. Finally we check the validity of Edwards' hypothesis for the 
athermal tapping dynamics. 
\end{abstract}
%\pacs{05.20.-y}{Classical statistical mechanics}
%\pacs{45.70.Cc}{Static sandpiles; granular compaction}
%\pacs{75.10.Nr}{Spin-glass and other random models}

\section{Introduction}

Granular matter and glasses share a number of properties -- such as 
off-equilibrium dynamics, aging, and hysteresis -- and analogies between them 
have long \cite{first} been pointed out. However it was
not until the seminal experiments of the Chicago group \cite{nagel}
on granular compaction were carried out that serious 
attempts were made to quantify such analogies. 
The experiments focus on the compaction behaviour of a large number of 
grains subject to repeated tapping and have become a paradigm for subsequent 
theoretical models. 

These models fall into roughly two classes: lattice-based
models \cite{amgcb,tetris,paj}
in a finite-dimensional space (which in general do not admit analytic 
solutions) or mean field models \cite{jorge,cugliandolo} (where each 
site interacts with a large number of other sites). In this paper we discuss 
how models on random graphs may be used to describe 
aspects of the behaviour of granular matter which 
depend on the {\em finite} connectivity of the (disordered) grains,
while still remaining analytically accessible.

The aim of this paper is twofold: We discuss and motivate a simple 
spin-model defined on a random graph introduced as a model of granular 
compaction in \cite{bergmehtalett}. 
In this model the random close packing density reached 
asymptotically after a large number of taps is identified with a dynamic 
phase transition. Secondly, we discuss an athermal dynamics 
\cite{bergmehtalett,dean}, consisting 
of alternating periods of thermal dynamics at a certain temperature and 
quenches at zero temperature which take the system to a metastable state.   

Next, the compaction curve of the tapping process will be 
discussed and its two main features, the single particle relaxation 
threshold and the random close packing density (dynamical transition), 
will be analyzed in detail. We then present results of numerical 
simulations of 
tapping-amplitude cycling with a  discussion of
 hysteresis and the asymptotic state 
of the model. Finally we investigate the statistical mechanics of 
the blocked configurations and discuss the validity of the so-called 
Edwards measure \cite{jorge}. 

\section{Random graph models and granular media}

A random graph \cite{bollobas} consists of a set of nodes and bonds, 
with the bonds connecting each node at random to a finite number of 
others, thus from the point of view of connectivity 
appearing like a finite-dimensional structure. 
Each bond may link up two sites (a graph) or more (a so-called 
hypergraph). 

Formally, a random graph of $N$ nodes and average connectivity $c$ 
is constructed by considering all $N(N-1)/2$ possible bonds between 
the nodes and placing a bond on each of them with probability $c/N$. 
In other words the connectivity 
matrix $C_{ij}$ is sparse and has entries $1$ (bond present) and $0$ 
(no bond), which are independent and 
identically distributed variables with probability $c/N$ and $1-c/N$ 
respectively. The resulting distribution of local connectivities is 
Poissonian with mean and variance $c$. 
The resulting structure is \emph{locally} tree-like but has loops 
of length of order $\ln(N)$. Although there is no geometric concept of 
distance (in a finite dimensional space), a chemical distance may be 
defined by determining the minimum number of steps it takes to go from one 
given point to another.  
  
\begin{figure}
\epsfysize=.25 \textwidth
  \epsffile{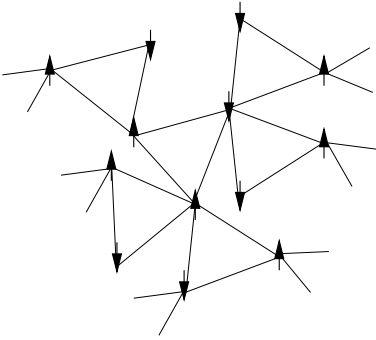}
\caption{A part of a random graph (strictly speaking a hypergraph) 
with triplets of sites forming plaquettes illustrating its locally 
tree-like nature (no planarity or geometric sense of distance are implied).}
\label{3graph}
\end{figure}

In a similar fashion, graphs -- strictly speaking hypergraphs -- 
with plaquettes connecting $3$ or more nodes 
each may be constructed. 
Choosing $C_{ijk}=1(0)$ randomly with probability 
$2c/N^2$ ($1-2c/N^2$) results in a random 3-hypergraph, where the number 
of plaquettes connected to a site is distributed with a 
Poisson distribution of average $c$. An illustration of part of 
such a graph is shown in Figure \ref{3graph}.  

Spin models on random graphs have been investigated for almost 20 years 
\cite{vianabray} since they may be considered as being halfway between 
infinite-connectivity models and finite-dimensional models, 
having to a certain 
extent the analytic accessibility of the former within the framework of 
mean-field theory, yet the finite connectivity of the latter. 
Interest in these models has intensified lately since they occur in the 
context of random combinatorial optimization problems \cite{remietal} 
and inroads have been made towards their analytic treatment beyond 
replica-symmetry \cite{mezpar,leoneetal}. 

In the context of modelling the compaction of granular matter, 
random graphs are the simplest structures with a finite number of neighbours. 
The finite connectivity is a key property, which goes
 beyond the simple fact that 
the grains in a granulate are in contact with a finite number of neighbouring 
grains. For instance kinetic constraints, which are a prominent feature 
in many models of granular behaviour \cite{kobandersen} can only be 
meaningfully defined in models with a finite connectivity. Furthermore 
cascades found experimentally during the compaction process may be 
explained by interactions between a finite number of neighbouring sites, 
where one local rearrangement sets off another one in its neighbourhood and so 
on \cite{bergmehtalett}. 

Another reason for the use of random graphs lies in the disordered structure 
of granular matter even at high densities. A random graph is the simplest 
object where a neighbourhood of each site may be defined, but has no  
global symmetries like a regular lattice. 
Additionally, the locally fluctuating 
connectivity may be thought of as modelling the range of 
coordination numbers of the grains \cite{pre}. 

\section{The model}

In the following we consider a 3-spin Hamiltonian on a 
random hypergraph where $N$ binary spins $S_i=\pm1$ interact in triplets 
\begin{equation}
\label{hdef}
H=-\rho N=-\sum_{i<j<k} C_{ijk} S_i S_j S_k 
\end{equation}
where the variable $C_{ijk}=1$ with $i<j<k$ denotes the presence 
of a plaquette connecting sites $i,j,k$ and $C_{ijk}=0$ 
denotes its absence.   

This Hamiltonian has recently been studied on a random graph 
in the context of satisfiability problems in combinatorial 
optimization \cite{hypersat}, on a random graph of fixed connectivity 
\cite{leoneetal}, and on a 2D triangular lattice 
\cite{newman,garrahan}. It has a trivial ground state where all spins 
point up and all plaquettes are in the configuration 
$+++$ giving a contribution of $-1$ to the energy. 
Yet, \emph{locally}, plaquettes 
of the type $--+,-+-,+--$ (satisfied plaquettes) also give the same 
contribution. This results in a competition between local and 
global satisfaction of the plaquettes: Locally any of the satisfied 
plaquettes are equivalent (thus favouring a paramagnetic state), 
yet globally a ferromagnetic state may be favoured, since there are 
few configurations satisfying all plaquettes, where $4$ configurations 
$+++,--+,-+-,+--$ occur in equal proportions.    
For $c>c_c \sim 2.75$ \cite{hypersat},  ground states have a positive 
magnetisation, which may be interpreted as the onset long-range order and of a 
possibly crystalline \cite{transunpub,jpcm} state of the granular medium. 

\begin{figure}
\epsfysize=.25 \textwidth
  \epsffile{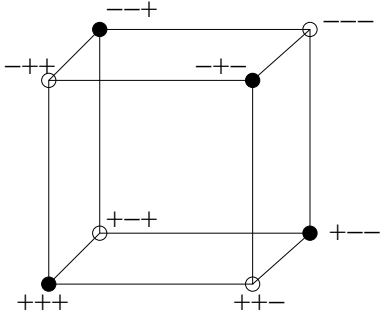}
\caption{The phase space of three spins connected by a single plaquette. 
Configurations of energy $-1$ (the plaquette is satisfied) are indicated 
by a black dot, those of energy $+1$ (the plaquette is unsatisfied) are 
indicated by a white dot.  
\label{figcube}}
\end{figure}

However \emph{two} spin flips 
are required to take a given plaquette from one satisfied configuration 
to another. Thus an energy barrier has to be crossed in any intermediate  
step between two satisfied configurations as illustrated in figure 
\ref{figcube}. In the context of 
granular matter this mechanism aims to model the situation where
compaction follows a temporary dilation; for example,
a grain could form an unstable ('loose') bridge with
other grains before it collapses into an available
void beneath the latter.
 This mechanism, by which an energy 
barrier has to be crossed in going from one metastable state to 
another, has recently been argued to be an important ingredient in models 
of granular compaction \cite{jpcm}. 

The crucial feature of the model responsible for the slow dynamics, 
however, is the degeneracy of the four configurations of plaquettes with 
$s_i s_j s_k=1$ resulting in competition between satisfying plaquettes 
 \emph{locally}  and \emph{globally}.
In the former case,
all states with even parity may be used, resulting in a large 
entropy and in the latter, only the $+++$ state may be used. 
A dynamics based on local quantities will thus \emph{fail} to find the 
magnetised configurations of low energy. 

This mechanism has a suggestive analogy in the concept of geometrical 
frustration of granular matter, if we think of plaquettes
as granular clusters. When grains are shaken, 
they rearrange locally, but locally dense configurations can be mutually
incompatible. Voids may appear between densely packed clusters
due to mutually incompatible grain orientations between neighbouring
clusters. The process of compaction in granular media consists
of a competition between the compaction of \emph{local} clusters
and the minimisation of voids \emph{globally}. 

\subsection{Modelling tapping}
\label{modellingtapping}

There have been many kinds of dynamical schemes to model
the behaviour of granular media under tapping. A recurring theme is the 
alternation of periods of randomly perturbing the system and 
periods in which the system is allowed to settle into a 
mechanically stable state. 
These have included nonsequential
Monte Carlo reorganisation schemes \cite{prl}, the ratio of upward 
to downward mobility of particles on a lattice \cite{tetris}, or 
variable rates of absorption and desorption \cite{prados1}. 

In the same spirit, we treat each tap as consisting of two phases. 
First, during the {\em dilation}
phase, particles are accelerated and are relatively
free to move with respect to each other for a time. 
In the second phase, the {\em quench} phase, particles relax until 
a mechanically stable configuration is reached. 

As initial condition we use a configuration obtained by quenching the 
system from a configuration where the spins are chosen independently to be
 $\pm1$ 
with equal probabilities. To mimic the action of tapping, 
we choose the following dynamics of the spins. The dilation phase is 
modelled by a single sequential Monte-Carlo-sweep of the system at a 
dimensionless temperature $T$: A site $i$ is chosen at random and flipped with 
probability $1$ if its spin $s_i$ is antiparallel to its local field $h_i$, 
 with probability $\exp(-h_i/T)$ if it is not, and with probability $0.5$ if 
$h_i=0$. This procedure is repeated $N$ times. 
Sites with a large absolute value of the local field $h_i$ 
thus have a low probability of flipping into the direction against the field. 
Such spins may be thought of as being highly constrained by their neighbours. 
This differs somewhat from the dilation-phase model in 
\cite{bergmehtalett,dean}, where a certain fraction of spins is 
flipped regardless of the value of their local field. We claim
that our present dynamics is rather more realistic in the context
of vibrated granular media; if grains are densely packed (strongly
'bonded' to their neighbours), they are less likely to be displaced
during the dilation phase of vibration than grains
which are loosely packed. 

The {\em quench} phase is modelled by a quench of the system at $T=0$, which 
lasts until the system has reached a blocked configuration, i.e. each site 
$i$ has $s_i=\mbox{sgn}(h_i)$ or $h_i=0$. Thus at the end of each tap the 
system will be in a blocked configuration.

This dynamics may be thought of 
as a series of quenches where the initial condition for each quench is 
obtained by perturbing the result of the previous quench.
It is a simplified version, suitable for spin models, of the 
tapping dynamics used in cooperative Monte Carlo simulations
of sphere shaking \cite{prl}. In the context of combinatorial 
optimization it corresponds to the class of random-restart algorithms 
(e.g. \cite{walksat}), which include some of the most efficient 
algorithms for the solution of optimization problems.

\section{The compaction curve}

\begin{figure}
\epsfysize=.5 \textwidth
  \epsffile{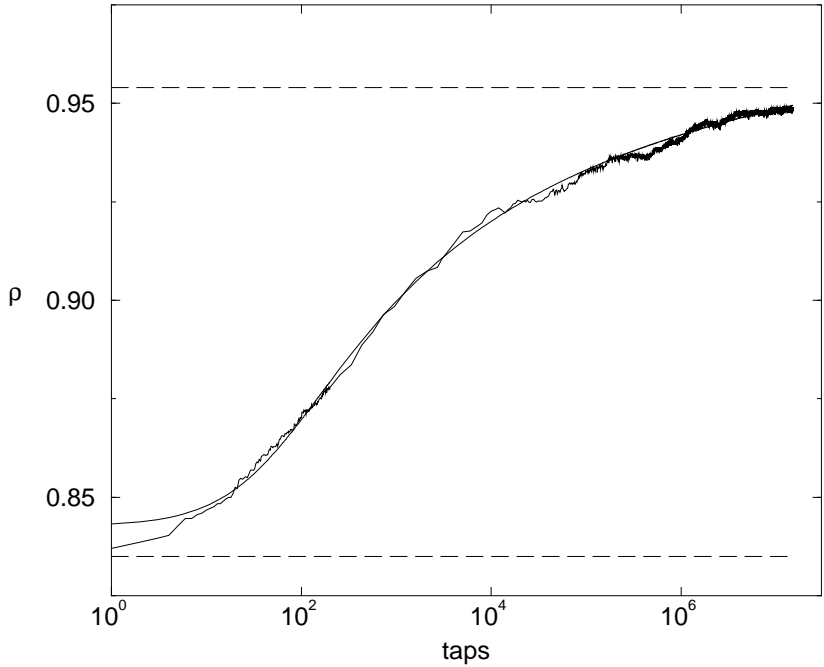}
\caption{Compaction curve at connectivity $c=3$ for a system 
of $10^4$ spins with $T=.4$. 
The data stem from a single run and 
the fit (smooth solid line line) follows (\ref{loglaw}) 
with parameters $\rho_{\infty}=.989$, $\rho_0=.843$, $D=4.716$, 
and $\tau=52.46$. The long-dashed line (top) indicates the approximate 
density $0.954$ at which the dynamical transition 
occurs, the long-dashed line (bottom) indicates the approximate 
density $0.835$ at which the fast dynamics stops, the 
{\it single-particle relaxation threshold}.}
\label{compact}
\end{figure}

An example of a single run of the system is 
shown in figure \ref{compact}. 

We can identify three regimes of the dynamics: first, a very fast increase 
of the density up to a density $\rho_0$ during the first tap, 
then a slow compaction regime which takes the density up to $\rho_\infty$, 
and finally an asymptotic regime. 

In the first regime, all sites orient their spins in parallel with the local 
field acting on that site. 
This quench corresponds to a {\em fast} 
dynamics whereby 
\emph{single} particles \emph{locally} find the orientation maximizing the 
density leading to the density of $\rho_0$ 
\cite{kineticconstraints}. 
In \cite{bergmehtalett} this density was termed the 
{\it single-particle relaxation threshold} (SPRT). 

The second phase of the dynamics consists of removing some of the remaining 
frustrated plaquettes and gives a logarithmically slow compaction 
\cite{nagel,paj}
leading from a density $\rho_0$ to $\rho_\infty$. 
The resulting compaction curve may be fitted to the well-known logarithmic 
law \cite{nagel}
\begin{equation}
\label{loglaw}
\rho(t)=\rho_{\infty}-(\rho_{\infty}-\rho_0)/(1+1/D \, \ln(1+t/\tau)) \ ,
\end{equation} 
which may also be written in the simple form 
$1+t(\rho)/\tau = \exp{\{D \frac{\rho-\rho_0}{\rho_{\infty}-\rho} \}}\ , $
implying that the dynamics becomes slow (logarithmic) as soon as 
the density reaches $\rho_0$.  
In this regime, most spins have a nonzero local field acting on them, 
which keeps them fixed in a certain direction \cite{barratzecch}. The 
corresponding grains are firmly held in place by their neighbours.  
However, during the dilation phase some of them have  
their orientation altered, altering the local fields 
acting on their neighbouring grains by a finite amount,   
which could cause them to flip in turn.
The dynamics of the grains with zero local field may alter  
the local field of their neighbours, and induce a previously blocked grain 
to flip. In this way a \emph{cascade} \cite{bergmehtalett} of flips may ensue. 
 
With increasing density, free-energy barriers rise up causing 
the dynamics to slow down according to (\ref{loglaw}). The point where 
the height of these barriers scales with the system size marks a
{\em breaking of the ergodicity of the dynamics}, a break-up of the 
phase-space into a large number (scaling exponentially with the system size) 
of disconnected clusters, and a saturation of the 
compaction curve. For small driving amplitudes, we thus identify the 
asymptotic density (random close packing) with a {\em dynamical 
phase transition} \cite{cugliandolo,monasson,franzpar,aging}. 

In the following we will examine in detail the SPRT and the dynamical 
transition.  

\subsection{The single-particle relaxation threshold}
\label{sprt}

The first tap is
 modelled as a quench at zero temperature. At the end of
this,
each site is connected to 
more (or as many)  unfrustrated plaquettes than frustrated ones. The spin of 
any site where this is not the case would flip under the zero-temperature 
dynamics, turning frustrated plaquettes into unfrustrated ones.  
The question of the density reached after a quench from random starting 
conditions is highly non-trivial, since its resolution involves the 
basins of attraction of the zero-temperature dynamics. 

The problem may be illustrated 
by considering a single site $i$ connected to $2k_i$ other sites 
and subject to the local field  
$h_i=1/2 \sum_{jk} C_{ijk} s_j s_k$. For random initial conditions, 
the values of $l_i=h_i s_i$ are binomially distributed with a probability 
of $C^{k_i}_{(k_i-l_i)/2} (1/2)^{k_i}$ if $k_i-l_i$ is even and zero if 
it is odd. If $l_i<0$ zero-temperature dynamics 
will flip this spin, turn $l_i$ to $-l_i$ and turn $(k_i\pm l_i)/2$ satisfied  
(dissatisfied) plaquettes connected to it into dissatisfied 
(satisfied) ones. This will cause the $l_j$ of $k_i\pm l_i$ neighbouring 
sites to decrease (increase) by 2. This dynamics stops when all sites 
have $l \geq 0$, giving $\rho_0=1/(3N) \sum_i l_i$. 

This process is made complicated by correlations between the local fields 
of neighbouring sites. Neglecting these correlations we arrive at a simple 
population model of $N$ units, each with a Poisson 
distributed value of $k_i$ and a value of $l_i$ distributed according to 
the initial binomial distribution. At each step a randomly chosen 
element with negative $l_i$ has its $l_i$ inverted, and 
$k_i\pm l_i$ randomly chosen elements have their values of $l$ decreased 
(increased) by 2 until $l_i \geq 0 \ \forall i$. 
This simplistic model works surprisingly well at low values of the 
connectivity $c$ (with an error of about $10\%$ up to $c=6$), but obviously 
fails completely at large values of $c$ or in fully connected models. 

In principle the differential equations describing the population dynamics 
could be solved analytically. Here we simply report the results for 
running the population dynamics numerically with $N=10^4$ at $c=3$. We obtain 
$\rho_0=0.835(1)$ which 
is shown as a dotted line in Figure \ref{compact}. Note that this 
density is found to be much higher than that of a typical 
'blocked' configuration 
with $l_i \geq 0 \ \forall i$ which is found to be $0.49$ 
(see section \ref{blockedconfiguration} and also the discussion in 
\cite{barratkurchan}).  
Despite the fact that these 'blocked' configurations
are exponentially dominant, the total basin of attraction of the 
configurations at $\rho_0$ dominates the space of random initial 
conditions. 
 
Another significant feature of this regime is that
a fraction of spins is left with local fields exactly equal 
to zero, which thus keep changing orientation \cite{barratzecch}. 
These spins may be compared to so-called rattlers, \cite{weeks}, 
i.e. grains which change their orientation {\em within} well-defined 
clusters \cite{pre}. These  will be used as a 
tool to probe the statistics of blocked configurations in section 
\ref{blockedconfiguration}. 

To conclude this section, the SPRT density appears as the 
density which is reached dynamically by putting each particle 
into its \emph{locally} optimal configuration, as has also been found in 
lattice-based models \cite{paj} and simulations of sphere packings 
\cite{jpcm,transunpub}, which show both fast and slow dynamics.

\subsection{The dynamic transition}
\label{dyntrans}

The dynamical transition is marked by the appearance of an exponential 
number of valleys in the free-energy landscape and thus a breaking of 
ergodicity \cite{monasson,franzpar,aging}.  
In the event that the dynamics is thermal, equilibration times diverge at the 
temperature corresponding to the dynamic transition. Cooling the 
system down gradually from high temperatures will also result in 
the system falling out of equilibrium at the dynamical transition 
temperature. Furthermore, the energy will get stuck at the energy at 
which the transition occurs. 

Since this phenomenon is the result of the drastic change in the 
geometry of phase space, it is not surprising that we also find it 
in the athermal dynamics defined in section \ref{modellingtapping}. 
Either gradually decreasing the tapping amplitude $T$ or   
tapping at a low amplitude for a long time will get the system to 
approach the density (energy) at which the dynamical transition occurs. 
We thus identify the random close packing limit in this model with 
a dynamical phase transition. 

To support this picture, we give a simple approximation for the density 
$\rho_\infty$ at which the dynamical transition occurs. 
Using the replica-trick $\ln Z= \lim_{n \to 0} \partial_n Z^n$
\cite{MPV} and 
standard manipulations, we obtain for the average of the $n$-th power of the 
partition-function of the Hamiltonian (\ref{hdef}) averaged over the ensemble
 of 
random graphs 
\begin{eqnarray}
\langle \langle &&Z^n \rangle \rangle = 
  \prod_{\vec{\sigma}} \int_0^1 dc(\vec{ \sigma}) \exp \left\{-N \left(
       \sum_{\vec{ \sigma}}c(\vec{ \sigma}) \ln(c(\vec{ \sigma})) \right. \right.  \\
       &&\left.\left.+ c/3
       - c/3 \sum_{\vec{ \omega},\vec{ \tau},\vec{ \sigma}}
       c(\vec{ \omega})c(\vec{ \tau})c(\vec{ \sigma}) \exp \{ \beta \sum_a 
       \omega^a \tau^a \sigma^a  \}
       \right)\right\} \ , \nonumber
\end{eqnarray}
where $c(\vec{\sigma})$ is an order parameter function defined on the 
domain of the $2^n$ vectors $\sigma^a=\pm1$.

The general replica-symmetric ansatz is incorporated in 
$c(\vec{\sigma})=\int dh P(h) \frac{e^{\beta h \sum \sigma^a}}{(2 \cosh(\beta H))^n}$. Taking $P(h)=\delta(h)$ gives the paramagnetic solution, valid in the 
high-temperature phase, resulting in a free energy $f(\beta)$
\begin{equation}
\label{eqfree}
\beta f(\beta)=-c/3 \ln (\cosh \beta) + \ln (2) \ .
\end{equation}
To determine the temperature at which the dynamics transition occurs, a 
replica-symmetry breaking (RSB) ansatz is required \cite{monasson,franzpar}. 
A simple variational ansatz \cite{BMW,hypersat,WZ} implementing one step of 
RSB  given by  
\begin{equation}
c(\vec{ \sigma})=\prod_{b=1}^{n/m}  \left\{ \frac
       {\int dh^b G_{\Delta}(h^b) e^{\beta h^b \sum_{a=(b-1)m+1}^{bm}\sigma^a}}
       {\int dh^b G_{\Delta}(h^b)[2 \cosh(\beta h^b)]^m} \right\} \ ,
\end{equation}
where $G_\Delta(h)$ is a Gaussian with zero mean and variance $\Delta$, 
gives the free energy subject to the variational ansatz 
$f(\beta)=\mbox{extr}_{\Delta,m}f_1(\beta,\Delta,m)$ with  
\begin{eqnarray}
\beta&& f_1(\beta,\Delta,m)= \frac{\int Dz(\beta\sqrt{\Delta}z)
        [2\cosh(\beta\sqrt{\Delta}z)]^{m-1}
        \sinh(\beta\sqrt{\Delta}z) }
        {\int Dz [2\cosh(\beta\sqrt{\Delta}z)]^{m} } \nonumber \\
        &&-\frac{1-c}{m} \ln (\int Dz [2\cosh(\beta\sqrt{\Delta}z)]^{m}) 
        -c/(3m)\ln(\int \int \int Dz_1 Dz_2 Dz_3 \nonumber \\
        &&\left[8 \cosh(\beta\sqrt{\Delta}z_1) \cosh(\beta\sqrt{\Delta}z_2) 
        \cosh(\beta\sqrt{\Delta}z_3)\cosh(\beta) + 
         8 \sinh(\beta\sqrt{\Delta}z_1) \sinh(\beta\sqrt{\Delta}z_2) 
        \sinh(\beta\sqrt{\Delta}z_3)\sinh(\beta) \right]^m ) \ , \nonumber
\end{eqnarray}
where $D(z)$ denotes the Gaussian measure with zero mean and
variance one. The dynamical transition occurs at a temperature 
\cite{footnttemp} where  
$\partial(\beta f(\beta,\Delta,m))/\partial m$ evaluated at $m=1$ 
develops a minimum at finite $\Delta$ \cite{monasson,franzpar}. The 
corresponding density is marked with a horizontal line in Figure 
\ref{compact} and agrees well with the asymptotic density reached by 
the tapping dynamics. However this asympotitic density is not the 
highest density, that can be reached without putting the system 
into an ordered configuration, however it is the highest such density 
which is reached by a local dynamics.  

\section{Amplitude cycling and the stationary state}

The tapping dynamics introduced in section \ref{modellingtapping} 
may be used to increase and decrease the tapping amplitude successively. 
This \emph{amplitude cycling} is an important protocol in real and  
numerical experiments. The ramp rate \cite{nagel} is defined as
the ratio $\delta(Amplitude)/\tau$ where $\tau$ is the number
of taps spent at each amplitude, which is changed with an increment of
$\delta(Amplitude)$ after each series of taps.

\begin{figure}[htb]
 \epsfysize=.5 \textwidth
  \epsffile{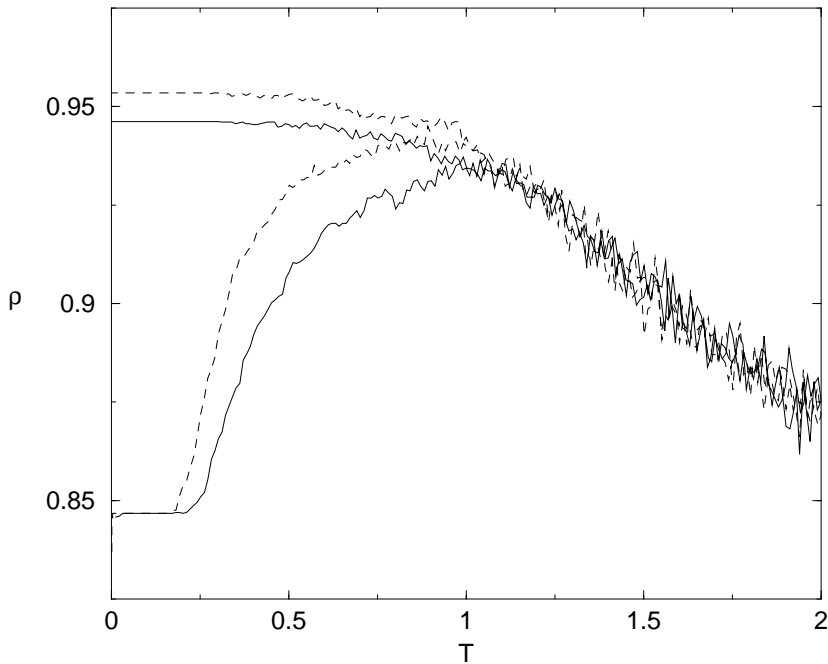}
\caption{The results of ramping the amplitude up and down again at 
two different rates, $10^{-4}$ (solid line) per tap and $10^{-5}$ (dashed 
line). The lower rate results in a steeper increase of the density 
during the increase of the amplitude (lower branches of solid and dashed 
lines). 
\label{figcycle1}
}
\end{figure}

The results of increasing $T$ continuously from $0$ to $2$ and 
back again at two rates $10^{-4}$ and $10^{-5}$ per tap is shown in figure 
\ref{figcycle1}; here, clearly $\tau = 1$, and the ramp rate
is in fact just $\delta(Amplitude)$. 
As expected, both at high and at low cycling rates, the 
density first reaches the SPRT $\rho_0$, then increases 
with increasing amplitude and time, until it decreases again at 
large values of $T$. As the amplitude is decreased, the system reaches 
$\rho_{\infty}$. The part of the curve where $T$ is increased for the first 
time is conventionally \cite{nagel} called the \emph{irreversible branch}, 
while the \emph{reversible branch} refers to the section where
$T$ is subsequently decreased and then increased again.

However the similarity of this simple picture with experimental results of 
\cite{nagel} is deceptive. From section \ref{dyntrans} we know that at 
fixed, finite but low amplitudes the model eventually reaches 
$\rho_{\infty}$. As a result, the branches of increasing amplitude at 
low $T$ do not coincide for high and low rates of change of 
the amplitude. At low rates of change the density as a function of $T$ is 
higher than at high rates of change. 
This irreversibility of the so-called reversible branch is thus 
due to the system not reaching a steady-state at each value of the amplitude. 
This behaviour has also been observed in other models 
\cite{coniglio,paj}.

In fact, in the limit of infinitely slow increments 
of $T$, the irreversible branch would disappear, 
and $\rho$ would become a 
single-valued function of $T$. This would be in direct
contradiction to the experimental results 
of \cite{nagel}, where at low amplitudes the steady-state 
density \emph{cannot} be reached with a sufficiently large number of taps,
at least within experimentally realisable times. 

This phenomenon may be thought of as follows: Some particles in a granular 
assembly are so strongly constrained that they will never 
(at least within experimentally realisable times) be moved by taps of a 
sufficiently low amplitude. In the dynamics of our model described
thus far, however, sites with a high local field 
may be flipped at any finite value of $T$ with a correspondingly 
\emph{small but finite} probability, leading the system eventually to 
$\rho_{\infty}$. 

To model this effect, it is not sufficient to prevent spins with a large 
value of the local magnetic field from being flipped 
since the dynamics of their neighbours will 
eventually lead to a reduction of their field, freeing the previously 
constrained spins.  Instead, we assign to each site $i$ a real number $r_i$ 
between zero and one, and during the dilation phase of the dynamics, we 
only flip spins at sites $i$ with $r_i<T$. During the 'quench' phase any  
spin may be flipped.  

In the language of grains, $r_i$ represents the strength with which
a grain is constrained by its neighbours; sites $i$ such that $r_i>T$
will be permanently resistant to being displaced at an intensity of
vibration $T$. In a real system, these thresholds would be 
determined by details of the inter-grain force network.  

The aim of this modification is to check if the scenario of section 
\ref{dyntrans} survives, since in principle it is possible that a new and 
lower value of $\rho_\infty$ emerges after the amplitude has been 
increased and decreased: 
At high shaking amplitudes frustrated plaquettes (clusters) might 
be generated, which cannot be eliminated at low values of the amplitude, 
similar to the scenario proposed in \cite{prados2}. 

\begin{figure}[htb]
 \epsfysize=.5 \textwidth
  \epsffile{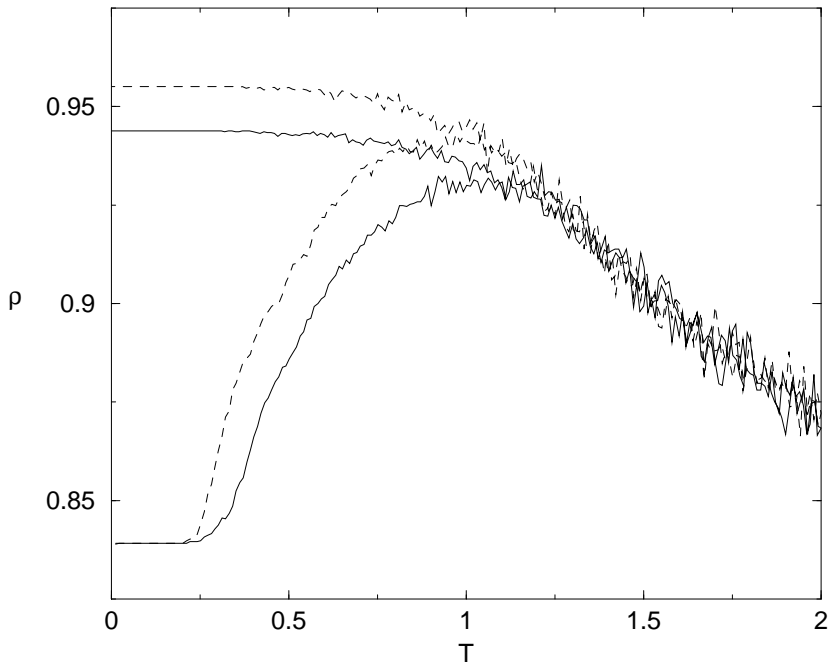}
\caption{The results of ramping the amplitude up and down again at 
two different rates, $10^{-4}$ (solid line) per tap and $10^{-5}$ (dashed 
line). This time a subset of the sites depending on the amplitude 
are constrained 'by hand' and do not flip. 
\label{figcycle2}
}
\end{figure}

\begin{figure}[htb]
 \epsfysize=.5 \textwidth
  \epsffile{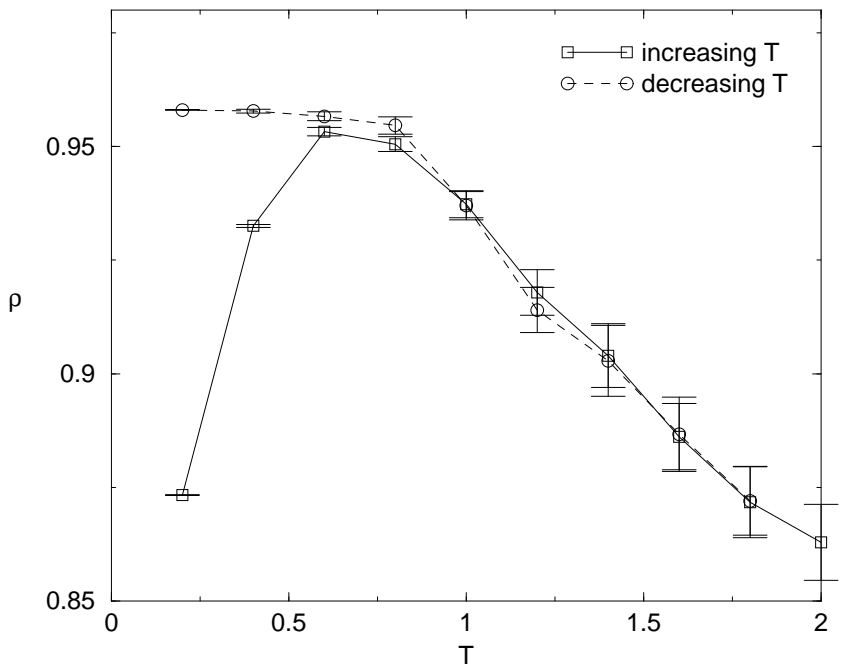}
\caption{The asymptotic density for tapping amplitudes ranging from $T=0.2$ to 
$T=2$ in steps of $0.2$. The density measured after $10^7$ taps at each 
amplitude and convergence to a steady-state each time was checked.  
\label{asympt} 
}
\end{figure}

Figure \ref{figcycle2}, where the cycling of Figure \ref{figcycle1} 
is repeated with 
the constraining of spins, shows that $\rho_\infty$ remains unaltered. 
In fact, Figures \ref{figcycle1} and \ref{figcycle2}
look remarkably similar. The effect of constraining some 
spins at low values of $T$ emerges when the ramp rate is decreased 
substantially, in particular by allowing the spins to 
'equilibrate' at each amplitude by choosing a large $\tau$. 
In figure 
\ref{asympt} we thus increase the shaking amplitude $T$ from 
$0.2$ to $2$ in steps of $0.2$ with $\tau = 10^7$ taps
waited at each amplitude step. This makes sure that a steady-state of the 
density has been reached at each amplitude. We find, in this case, 
that at low amplitudes the immobile spins cause the system to 
reach a steady-state with a density lower than $\rho_\infty$. Despite
the fact that in \cite{coniglio,paj}, very low ramp
rates were used, with large 'waiting times' $\tau$ at each tap,
this behaviour was not observed; rather the results of all these
simulations implied that the asymptotic density  $\rho_\infty$ would
always be approached in the limit of sufficiently low ramp rates.

One may 
view the (random) configuration of the immobile spins at each value of 
$T$ as an additional quenched disorder and their effect on neighbouring 
mobile spins, as a random local field. Presumably the dynamics in the 
sub-space of phase-space corresponding to the mobile spins (with fixed 
local fields due to the immobile spins) undergoes a dynamic 
transition as the corresponding steady-state density is reached. 
The result that $\rho_\infty$ is reached decreasing $T$ from above 
even though a finite fraction of spins has been rendered immobile at 
low temperatures, is quite remarkable: it is a testament to the paramagnetic 
nature of the model at densities below $\rho_\infty$. In a glassy state, 
one would expect the configurations of spins reached at high values of $T$ 
and subsequently frozen to alter the behaviour of the system at 
lower values of $T$. The paramagnetic system manages very well to adapt 
the \emph{mobile spins} to the configuration of the immobile ones produced at 
higher values of $T$ - not however to random configurations 
of the \emph{immobile spins}, which are responsible for the irreversible 
branch.

These results demonstrate a rather fundamental difference between
thermal excitations in glassy systems and intensities of mechanical vibration
in granular media. In the glassy phase of a system, one would expect 
the configurations of spins reached at high values of temperature
and subsequently frozen, to alter the behaviour of the system at 
lower values of temperature. In granular media, however, it is important
to let the system \emph{reach the asymptotic density} at each value of 
the shaking intensity$T$, in order even to begin to observe
the hysteresis that must result when mobile grains become part of immobile
clusters \cite{bmmb}, generating a type of 'quenched' disorder at least
at low vibrational intensities. The difference between Figs. \ref{figcycle2}
and \ref{asympt} clearly illustrates this. Given that the experiments
results \cite{nagel} were done in such a way that the system was allowed
to reach the asymptotic density at each value of the tapping amplitude,
our results indicate that 'jamming' \cite{nagel1} of grains
caused by the force network might be responsible for the fact that
the $\rho_\infty$ is not reached by tapping solely at low amplitudes.

\section{Blocked configurations and the Edwards-measure}
\label{blockedconfiguration}

In this section we focus on the statistical mechanics of the blocked
configurations referred to earlier, and use these results to address the 
question of ergodicity of the tapping dynamics.
After each tap according to section \ref{modellingtapping}, the system 
is in a blocked configuration, i.e. each site $i$ has $s_i=\mbox{sign}(h_i)$ 
or $h_i=0$. The Edwards hypothesis \cite{sam} 
states that in the steady state along the reversible branch 
all mechanically stable configurations 
at a given density are 
equiprobable. We test this hypothesis for the 
tapping dynamics of section \ref{modellingtapping}.  

\begin{figure}
 \epsfysize=.5 \textwidth
  \epsffile{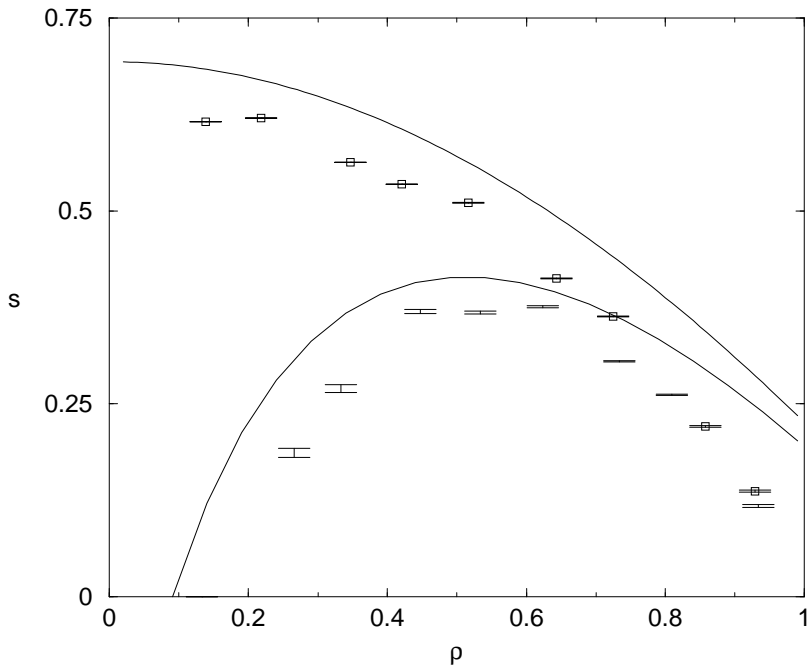}
\caption{The paramagnetic entropy (top) and the entropy of blocked states 
in the annealed approximation (bottom) for $c=3$ 
as a function of the density $\rho$. The data 
with errorbars shows the results of exhaustive enumerations of a system 
with $N=28$ averaged over $100$ samples. The results for the 
paramagnetic state are marked with squares.  
\label{figsbann} 
}
\end{figure}

We begin by calculating the average entropy of blocked configurations at 
a given density. In principle, we would need to average the logarithm of the 
number of blocked states over the ensemble of random graphs; this so-called 
quenched average can be expected to be self-averaging. For simplicity, 
we restrict ourselves to the so-called annealed average and compute 
\begin{equation}
\label{sbann}
s_{\mbox{\small{annealed}}}(\rho)=\frac{1}{N}\ln \langle\langle {\cal N}(\rho) \rangle\rangle\geq 
s_{\mbox{\small{quenched}}}(\rho)=\frac{1}{N} \langle\langle \ln \left({\cal N}(\rho)\right) \rangle\rangle \ ,
\end{equation} 
which gives an upper bound to the quenched average. The number of blocked 
configurations  ${\cal N}(\rho)$ may be written easily as 
\begin{equation}
\label{nblocked}
{\cal N}(\rho)=\prod_i \left( \sum_{s_i=\pm 1} \sum_{h_i=-\infty}^{\infty} 
        \delta ( h_i;1/2 \sum_{j,k} C_{ijk}s_j s_k )
        \Theta \left( h_i s_i \right) \right) 
        \delta \left( \rho - 1/(3N)\sum_i h_i s_i \right) \ ,
\end{equation}  
where $\delta(x;y)=1 \ \mbox{if} \  x=y$ and $0$ otherwise, denotes a 
Kronecker-delta and $\Theta(x)$ denotes a discrete Heaviside step function 
with $\Theta(x)=1 \ \mbox{if} \ x \geq 0$ and $0$ otherwise. 
After using integral representations for the Kronecker-deltas and 
standard manipulations, one easily obtains 
\begin{equation}
\label{sbannres}
s_{\mbox{\small{annealed}}}^{\mbox{\small{blocked}}}(\rho)=\mbox{extr}_{a,b,\beta} \left[ -\beta \rho 
        + 8c/3 \ (a^3+ b^3) -c/3 + \ln \left( 2 \sum_{h=1}^{\infty} 
(e^{\beta/3} a/b)^h I_h( 4 c a b ) 
        + 2 I_0( 4 c a b ) 
\right) \right] \ ,
\end{equation} 
where $I_h(x)$ denotes the modified Bessel-function of the first kind of 
order $h$. In Figure \ref{figsbann}, the entropy of blocked configurations 
$s_{\mbox{\small{annealed}}}(\rho)$ is shown along with the paramagnetic entropy given by (\ref{sannres}) below, which is derived by including terms 
with negative $s_i h_i$. 

These expressions were evaluated for $c=3$ and the results are shown in 
figure \ref{figsbann}.  
From this data, the lowest density at which blocked configurations occur 
($s_{\mbox{\small{annealed}}}(\rho)=0$) is found to be $\rho=.09$, and the density 
of a randomly chosen blocked configuration 
(the maximum of $s_{\mbox{\small{annealed}}}(\rho)$) is $\rho=.49$. 

Similarly, one may calculate the fraction $g$ of connected sites with 
zero local magnetic field in a blocked configuration. As we will argue 
below, this is a useful quantity to test the Edwards hypothesis. 
>From (\ref{sbannres}) one obtains  
\begin{equation}
\label{gbann}
g=\frac{I_0( 4 c a b ) }
{\sum_{h=1}^{\infty} (e^{\beta/3} a/b)^h I_h( 4 c a b ) + I_0( 4 c a b ) 
} - e^{-c} \ ,
\end{equation}  
where for any given value of $\rho$, the values of $\beta$ and of 
the order parameters $a$ and $b$ are given by the extremisation condition 
in equation (\ref{sbannres}), and where we have subtracted a trivial term $e^{-c}$ 
corresponding to the fraction of unconnected sites. 

Analogously, the fraction of connected sites with zero local magnetic field 
at a given density (without the blocking condition) is given by 
\begin{equation}
\label{gann}
g \prime =\frac{I_0( 4 c a b ) }
{\sum_{h=1}^{\infty} \left[ (e^{\beta/3} a/b)^h+(e^{-\beta/3} b/a)^h \right] 
        I_h( 4 c a b ) 
        + I_0( 4 c a b ) 
} - e^{-c} \ ,
\end{equation} 
where the values of $a,b, \mbox{and} \beta$ follow from extremizing over 
\begin{equation}
\label{sannres}
s_{\mbox{\small{annealed}}}(\rho)=\mbox{extr}_{a,b,\beta} \left[ -\beta \rho 
        + 8c/3 \  (a^3+ b^3) -c/3 +\ln \left( 2 \sum_{h=1}^{\infty} 
\left[ (e^{\beta/3} a/b)^h+(e^{-\beta/3} b/a)^h \right] I_h( 4 c a b ) 
        + 2 I_0( 4 c a b ) 
\right) \right] 
\end{equation}  

\begin{figure}[htb]
 \epsfysize=.5 \textwidth
  \epsffile{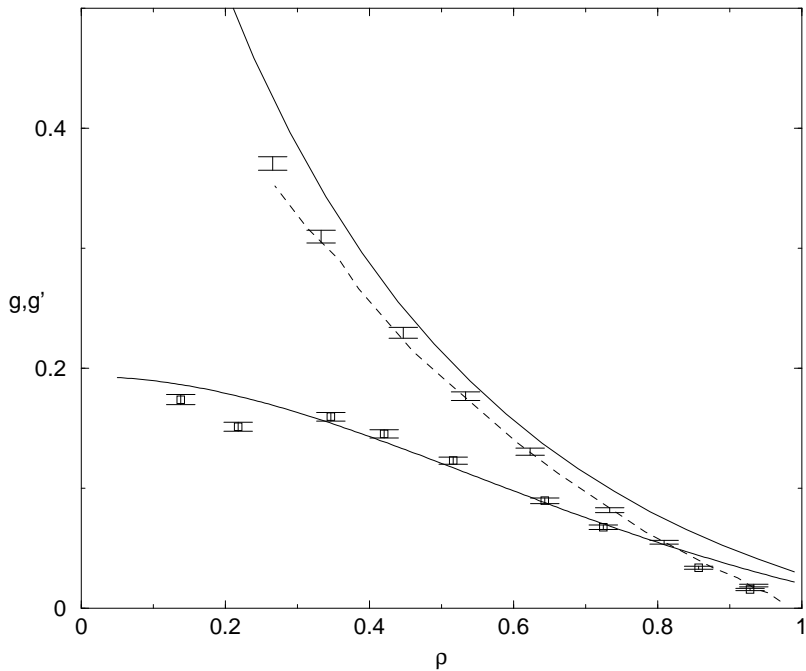}
\caption{The fraction of connected spins with zero magnetic field 
plotted against $\rho$ for $c=3$. 
The top solid line gives the analytical result $g$ (\ref{gbann}) for blocked 
configurations, the bottom solid line gives that for the paramagnetic state 
$g \prime$ (\ref{gann}) (without the blocking condition). 
The results of the quenched average are shown 
as a dashed line. The data 
with errorbars shows the results of exhaustive enumerations of a system 
with $N=28$ averaged over $100$ samples. The results for the 
paramagnetic state are marked with a square.  
\label{figg_rho} }
\end{figure}

Figure \ref{figg_rho} shows 
$g$ and $g \prime$ as a function of $\rho$. As expected, both quantities 
decrease monotonically with $\rho$. Again, we also give the results 
of exhaustive enumerations of a system with $N=28$, averaged 
over $100$ samples in order to test the validity of the annealed 
approximation. The annealed curve for blocked states and the numerical 
results show significant differences. For this reason, we also give the 
result of the much more involved replica-calculation of the quenched average 
\cite{berginprep} as the dashed line, which agrees very well with the 
numerical results. 

With these results, we may now test the Edwards hypothesis for this model 
under the tapping dynamics of section \ref{modellingtapping}. If blocked 
states at the asymptotic density are accessed with equal probability, a plot 
of the fraction of connected sites with zero local field versus the density  
should coincide with the results of (\ref{gbann}) at the asymptotic density. 

The value of $g$ is a useful  
quantity to test this hypothesis, since as discussed in section \ref{sprt}, 
the fraction of spins with a given local field is intimately related to 
the fast quenching dynamics. Also, 
as a one-time quantity it may be measured easily. 

\begin{figure}[htb]
 \epsfysize=.5 \textwidth
  \epsffile{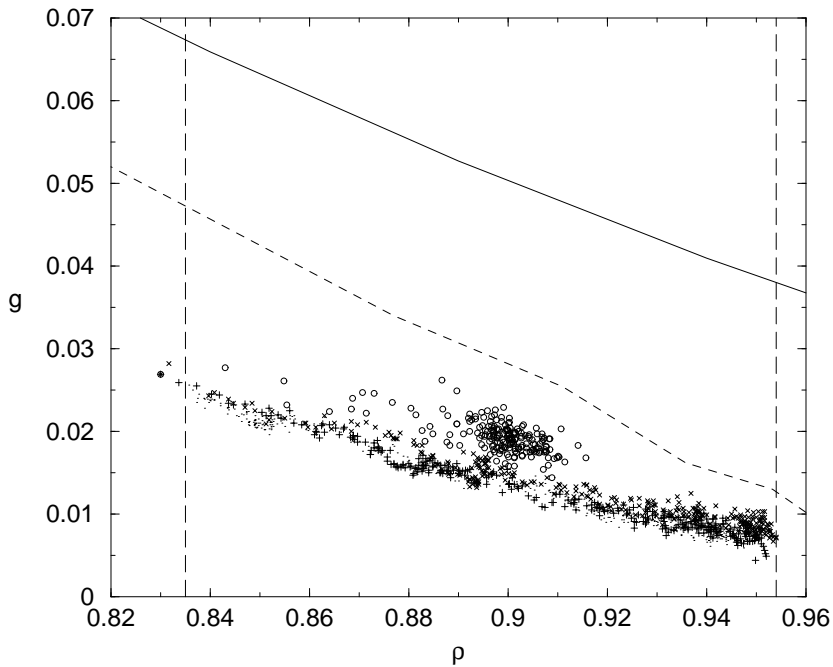}
\caption{The fraction $g$ of connected sites with zero local magnetic field 
during $4$ runs of a system with $N=10000, c=3$ at $T=0.4$ (dots), 
$T=0.56$ ($+$), $T=0.7$ ($\times$), and $T=1.5$ (circles). The solid line 
gives the analytic annealed 
result of (\ref{gbann}), the dashed line the corresponding quenched 
result. The lines left and right 
indicate the approximate values for $\rho_0$ and $\rho_{\infty}$ 
respectively.  
\label{figedwards_test} 
}
\end{figure}

In Figure \ref{figedwards_test}, we show the results of 4 single runs 
at $T=0.4,0.56,0.7,1.5$, plotting $g$ against $\rho$. It shows clearly 
that at the asymptotic density, the (quenched) result 
for $g$ against $\rho$ and the results of the tapping dynamics agree to within 
numerical accuracy of the analytical result, We tentatively conclude 
that the Edwards hypothesis is valid in this model at low tapping 
intensities and at the asymptotic density reached by low-intensity
tapping. Further results will be reported elsewhere \cite{berginprep}.   

During the compaction phase, however, the blocked configurations accessed 
dynamically have a lower value 
than that of the exponentially dominant blocked configurations contributing 
to (\ref{gbann}).
The result that the blocked configurations accessed dynamically have a lower 
value of $g$ than the typical ones at the same value of $\rho$, 
may be explained to a certain extent as follows.
Configurations with a large value of 
$g$ are favoured entropically, since spins with zero magnetic 
field may be flipped leaving the configuration blocked (provided 
this does not cause the local field of their neighbours to change sign).  
In the quenching dynamics, however, there is no such mechanism; sites 
with zero magnetic field are created only by spin-flips of neighbouring 
sites. 

Nevertheless it is remarkable that 
for the three lower values of $T$ (light tapping), where the asymptotic 
density is very close to $\rho_{\infty}$, the three traces nearly 
fall onto a single line, indicating that \emph{also during compaction} 
the blocked configurations 
are sampled according to a certain ensemble which depends on the density 
only. This occurs even though the plots of $rho$ versus the number 
of taps do not coincide for these amplitudes.  

\section{Conclusion}

To conclude, we have presented a \emph{finitely connected} 
spin model of vibrated granular matter, where we have built upon
earlier work \cite{bergmehtalett}. We argue that spin-models on random 
graphs may serve as models of granular matter, since they show no 
symmetries in the way a regular lattice does. They also arise as the 
Bethe-lattice approximation to finite-dimensional models. 
Multi-spin interactions 
generically arise when models of geometric frustrations are 
transferred to the Bethe-lattice. We discuss one of the simplest 
models of this class, the ferromagnetic $3$-spin model. Due to 
competition between satisfying the interactions globally and locally 
the model never reaches the ferromagnetic state. This mechanism 
aims to model the geometric frustration incurred by packings which arise 
from maximising the density locally. 

We also discuss the problems incurred by glassy models in the context 
of amplitude cycling.  
In order to model the effect of some grains being rendered completely 
immobile by large intergranular forces, we investigate the effect of 
constraining 'by hand' some of the spins 
in the context of amplitude cycling experiments. 
We also test the Edwards hypothesis \cite{sam} in the context
of our model, and concluded that it is valid for the tapping dynamics used. 

\bf{Acknowledgements:} It is a pleasure to thank  S. Franz, 
F. Ricci-Tersenghi, and R. Zecchinafor illuminating discussions.


\begin{thebibliography}{0}
\bibitem{kineticconstraints}{During the fast regime one may also model 
the mobility of the grains by a dynamics of the bonds subject to 
kinetic constraints. 
If bonds are allowed to move locally subject to kinetic constraints 
the resulting graph typically differs from those constructed according to 
(\ref{hdef}), however the compaction dynamics does not change 
qualitatively. }
\end{thebibliography}
\end{document}